\documentclass[twocolumn,showpacs,preprintnumbers]{revtex4}

\usepackage{graphicx}
\usepackage{dcolumn}
\usepackage{bm}

\begin{document}

\preprint{APS/123-QED}

\title{Effective Critical Exponents \\
of Ising Strips $D\times L$ with $(D\ll L)$ }
\author{M${{}^a}$ Felisa Mart\'\i nez}
\author{Carlos Garc\'\i a}
\author{Julio A. Gonzalo}
\email{julio.gonzalo@uam.es}
\affiliation{%
Departamento de F\'\i sica de Materiales, Universidad Aut\'onoma de Madrid\\
Cantoblanco, 28049 Madrid, Spain.
}
\date{\today}

\begin{abstract}
Monte Carlo data simulating phase transitions in Ising strips $D\times L,$ ($D\ll
L) $ with periodic boundary conditions show that $T_{c}(D)=0$ for $D\leq
D^{\ast }\simeq 6$ and $0<T_{c}(D)<T_{c}(d=2)$ for $D>D^{\ast }.$ Regular
scaling of $ML^{\beta /\nu }$ vs $|T-T_{c}|L^{1/\nu }$ is obtained only for $%
D>D^{\ast }$and the Monte Carlo effective susceptibility critical exponent $\gamma_{eff} (D)$
is shown to be well described by $\gamma (d)=\beta (d)[\delta (d)-1]$ with $%
d_{eff}(D)$ given by $d_{eff}(D)\simeq 1.5+(\frac{1}{200})(D-6)$ and $\beta (d)=(\frac{3d%
}{16}-\frac{1}{4}),$ $\delta ^{-1}(d)=(\frac{2d}{15}-\frac{1}{5})$, which can be understood
as valid with $d_{eff}(D)$.

\end{abstract}


\maketitle
\section{\label{sec:level1}Introduction}

It is well known \cite{1} that systems at criticality, for instance Ising
magnets at ($T\rightarrow T_{c},$ $H\rightarrow 0),$ are usually not only
scale invariant but also conformally invariant, i.e. the pertinent local
transformations look like combinations of dilations, rotations and
traslations, but shear distortions are not allowed. This fact has been used
to prove that in two dimensions, the Ising critical exponents must be
fractional.

Monte Carlo methods \cite{2} have been extensively used to characterize
phase transitions in Ising systems and are used in this work to stidy $D\times L$ strips.

\begin{figure}
\includegraphics[width=6.1cm,height=7.9cm,angle=270]{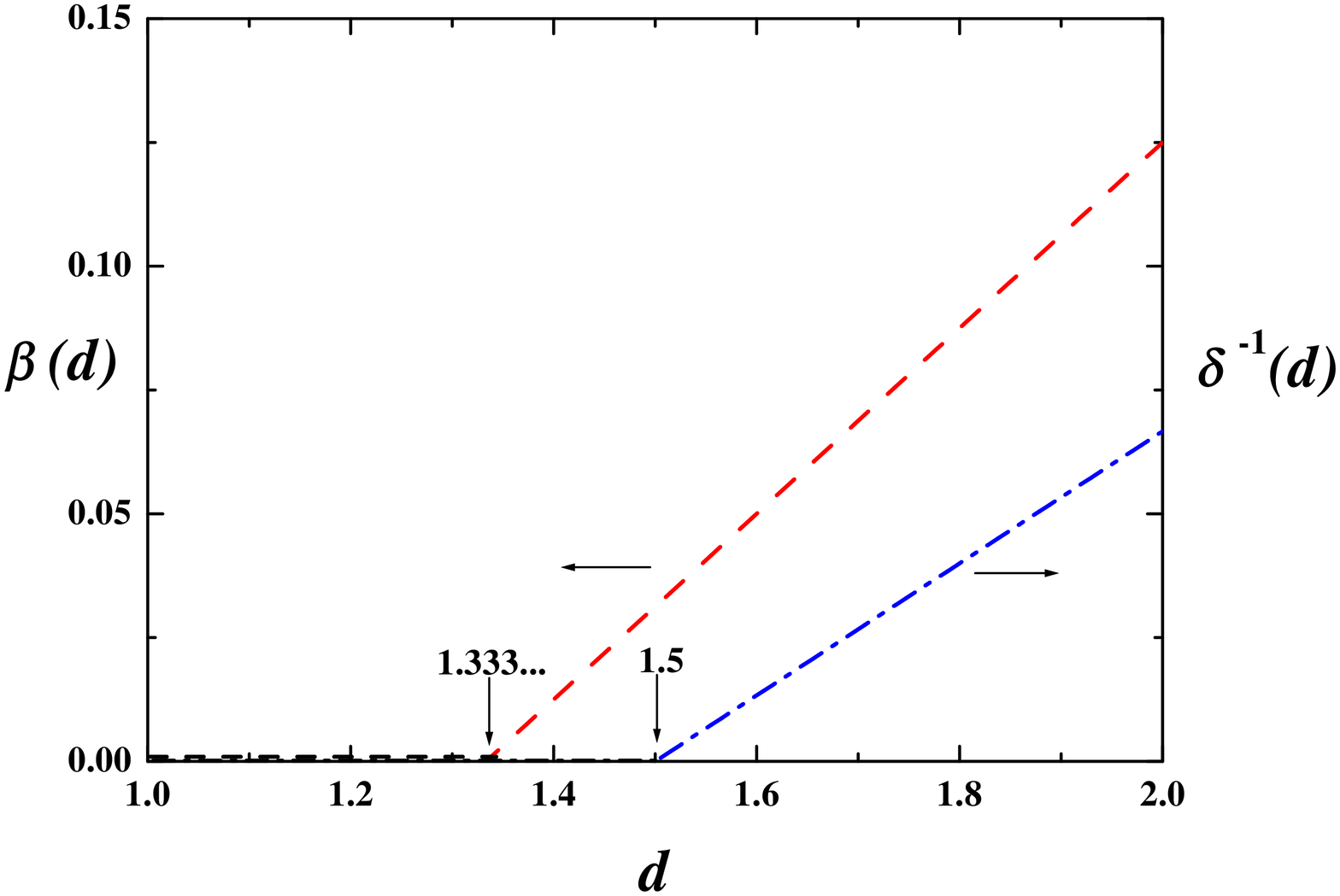}
\\
\caption{\label{fig:Fig1}Critical exponents $\protect\beta (d)=(\frac{3d}{16}-\frac{1}{4})$
and $\protect\delta ^{-1}(d)=(\frac{2d}{15}-\frac{1}{5})$ extrapolated from
the respective fractional values $[\protect\beta (4)=1/2$, $\protect\beta %
(2)=1/8]$ and $[\protect\delta ^{-1}(4)=1/3$, $\protect\delta ^{-1}(2)=1/15]$%
.}

\end{figure}

Recent Monte Carlo \cite{3} data on large three
dimensional Ising lattices ($L^{3}\simeq 1.5\times 10^{6}$ spins) appear to
indicate that for $2\leq d\leq 4$ (including $d=3)$ the critical exponents
are either well defined fractional \cite{4} or indistinguishable from fractional, and that they are
determinated by compact expressions for $\beta (d)$ and $\delta ^{-1}(d).$
(See Figure\ref{fig:Fig1})

%
%

In the present work we carry out an empirical investigation of the critical
behaviour of strips $D\times L$ (with increasing width $D\ll L)$ by means of
Monte Carlo simulations of the phase transition for long strips with $%
D=2,3,4...$ up to $D=52$, using periodic boundary conditions, to ascertain
whether the effective critical exponents $\beta (d)$ and $\delta ^{-1}(d)$ describing
the phase transition are given by the same expressions linear in $d$ which
describe well the behaviour for higher dimensionalities ($d=2,3,4)$.

\section{\label{sec:level1}Monte Carlo results}

Numerical finite size simulations of the phase transitions in $D\times L$
Ising systems with periodic boundary conditions were performed for $D=2$ to $%
D=52$. Wolff cluster algorithms \cite{5} were
used in strips of length $250\leq L\leq 5000$. Periodic boundary conditions
were used always in the $L$ direction but they had not much effect for large $%
L$ values as it is to be expected. The thermalization time, relaxation time and number of states
were increased steadily until the final results were not appreciably affected by further
increases. $140000$ Monte Carlo steps per spin for each temperature were
taken. To reduce the critical slowing down at $T\simeq T_{c}$ a single
cluster Wolff algorithm was used. Initial conditions at a any given
temperature in a closely spaced set of temperatures ($\Delta T\leq 0.01$) were
taken from the equilibrium conditions at the previous temperature.

The critical temperature of strips with width $D\ll L$ which is a non-universal quantity, was
determined by means of the Binder Cumulant method \cite{2}, crossing
data for $D\times L $ with data for $D\times 2L$. It was found that for $%
D\leq 6$ no crossing took place at $T>0$, and it was checked whether scaling
of $ML^{\beta /\nu }$ vs $|\epsilon |L^{1/\nu }$, with $|\epsilon |=|T-T_{c}|
$, was possible using $\beta =0$ and $\nu \leq 1/2$. It was found that no
such scaling took place, but, what could be called $\ "one-dimensional"$
scaling \cite{1}, with $|\epsilon |=e^{-4/T}$ was observed to take place for 
$D=2,3,4,5$.

For $D>6$ the Binder Cumulant method did provide non-cero transition
temperatures given by \cite{2}

\begin{equation}  \label{eq:1}
T_{c}(D)=T_{c}(d=2)[1-e^{-m\sqrt{D-D^{\ast }}}]
\end{equation}

with $D^{\ast }\simeq 6,$ and $m\simeq 0.353\pm 0.011$; and regular scaling
of $ML^{\beta /\nu }$ vs $|\epsilon |L^{1/\nu }$, with $|\epsilon
|=|T-T_{c}| $ was observed to hold, with $\beta (d)$ and $\nu (d)$ evolving
smoothly between the respective values for $d=1.5$ and $d\simeq 2$ as
specified below.

To describe the evolution of the critical exponents as a function of
two-dimensional strip width $D$ we note first that $\beta (d)$, $\delta
^{-1}(d)$ and $\gamma (d)=\beta (d)[\delta (d)-1]$ for $d=2,$ and $d=4$ are
given by fractional values specified by

\begin{equation}  \label{eq:2}
\beta (d)=\left( \frac{3d}{16}-\frac{1}{4}\right)
\end{equation}

\begin{equation}  \label{eq:3}
\delta ^{-1}(d)=\left( \frac{2d}{15}-\frac{1}{5}\right)
\end{equation}

\begin{equation}  \label{eq:4}
\gamma (d)=\beta (d)[\delta (d)-1]= \newline \frac{11}{8}+\frac{45}{192}\left( \frac{2%
}{2d-3}\right) -\frac{3}{16}\left( \frac{2d-3}{2}\right)
\end{equation}

For $d=3$ Monte Carlo data \cite{3} are consistent within narrow statistical
error bars with fractional exponents $\beta (d=3)=\frac{5}{16},$ $\delta
^{-1}(d=3)=\frac{1}{5}$ and $\gamma (d=3)=\frac{5}{4}$, as given by
Equations (\ref{eq:2}), (\ref{eq:3}) and (\ref{eq:4}) respectively.

\begin{figure}
\includegraphics[width=6.1cm,height=7.9cm,angle=270]{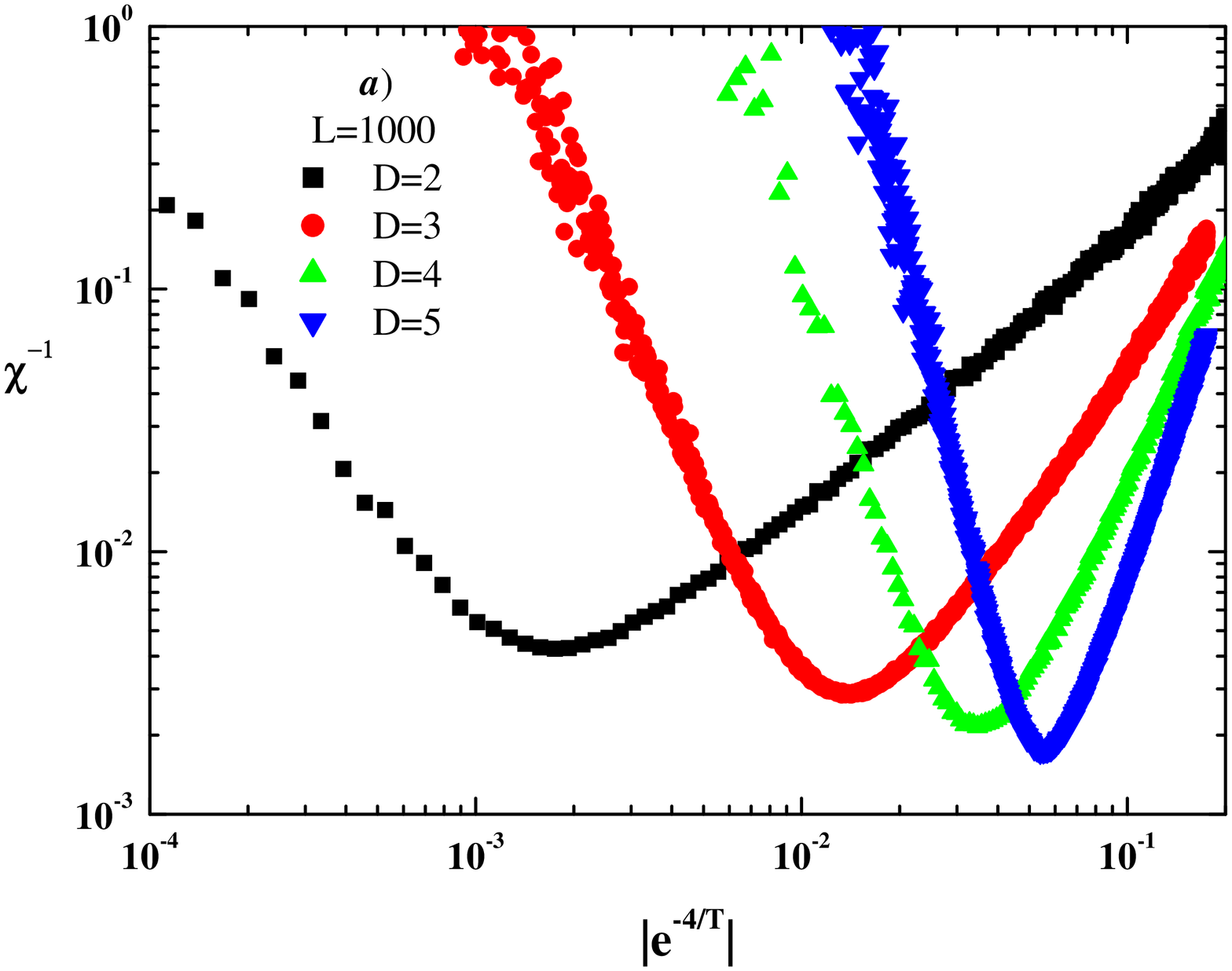}
\\
\caption{\label{fig:Fig2}Inverse susceptibility vs $|\protect\epsilon |=|e^{-4/T}|$ for $%
L\times D$ strips with $L=1000$ and $D=2,3,4,5$ showing evolution of $%
\protect\gamma_{eff} (D)$ for $D\leq D^{\ast }\simeq 6$.}
\end{figure}

\begin{figure}
\includegraphics[width=6.1cm,height=7.9cm,angle=270]{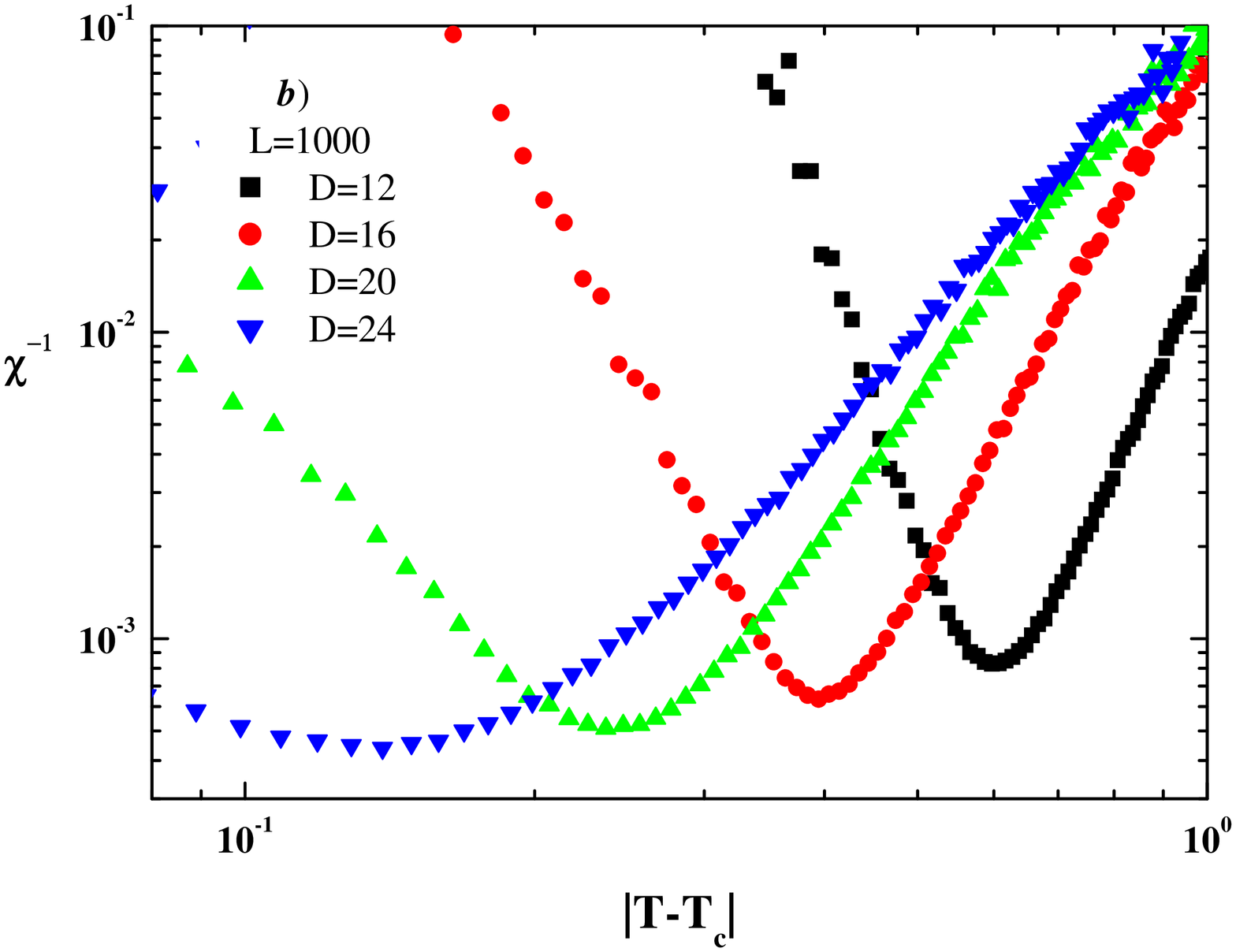}
\\
\caption{\label{fig:Fig3}Inverse susceptibility
vs $|\protect\epsilon |=|T-T_{c}|$ for $L\times D$ strips with $L=1000$ and $%
D=12,16,20,24$; showing evolution of $\protect\gamma_{eff} (D)$ for $D\geq D^{\ast
}\simeq 6$.}

\end{figure}

Figure \ref{fig:Fig2} shows plots of $\chi ^{-1}$ vs $|\epsilon
|=|e^{-4/T}|$ for $D<D^{\ast }=6$ and Figure \ref{fig:Fig3} shows like plots of $\chi ^{-1}$
vs $|\epsilon |=|T-T_{c}|$ for $D\geq D^{\ast }$. It can be seen
that $\gamma_{eff} (D)$ diverges at a width $D\simeq 6$ from both sides. From these
data the evolution of $\gamma_{eff} (D)$ can be determinated with fair accuracy,
and it can be used in conjunction with Equation (\ref{eq:4}) to determine
the effective dimensionality of strips with growing $D$.

\begin{figure}
\includegraphics[width=6.1cm,height=7.9cm,angle=270]{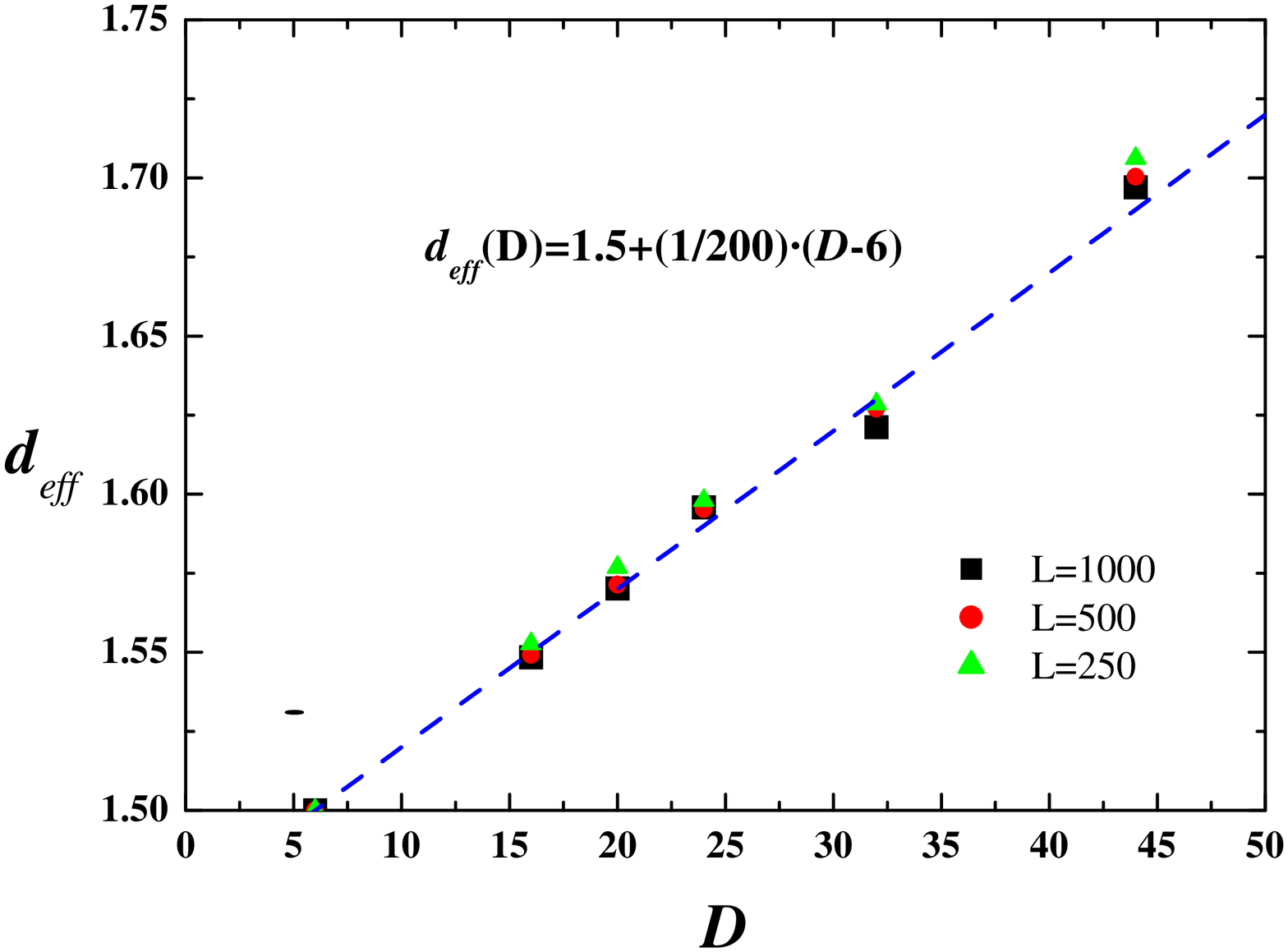}
\\
\caption{\label{fig:Fig4}Effective dimensionality $d$ of $L\times D$ strips as a function of
$D$ for $D\geq D^{\ast }\simeq 6$ obteined by means of the relationship $%
\protect\gamma_{eff} (D)=\protect\gamma (d)=\protect\beta (d)[\protect\delta
(d)-1] $ as given by means of linear extrapolations of $\protect\beta (d)$
and $\protect\delta ^{-1}(d)$ from the known fractional values at $d=2$ and $%
d=4$ (see text).}

\end{figure}

This is done in Figure \ref{fig:Fig4} where a plot of $d_{eff}(D)$ results in a
quiasi-linear dependence of $d_{eff}(D)$ starting at $d_{eff}(D^{\ast })=1.5$
(corresponding to $\delta ^{-1}(d)=0$ after Equation \ref{eq:3}) and
growing up from this value towards $d_{eff}(L)=2$, i.e. for rectangular $L\times D$
strips with $D\rightarrow L$ (approaching square lattices). $d_{eff}(D)$ therefore, is approximately given by

\begin{equation}  \label{eq:5}
d_{eff}(D)=1.5+\left( \frac{1}{200}\right) \left( D-D^{\ast }\right)
\end{equation}

where $D^{\ast }\simeq 6$.

\begin{figure}
\includegraphics[width=6.1cm,height=7.9cm,angle=270]{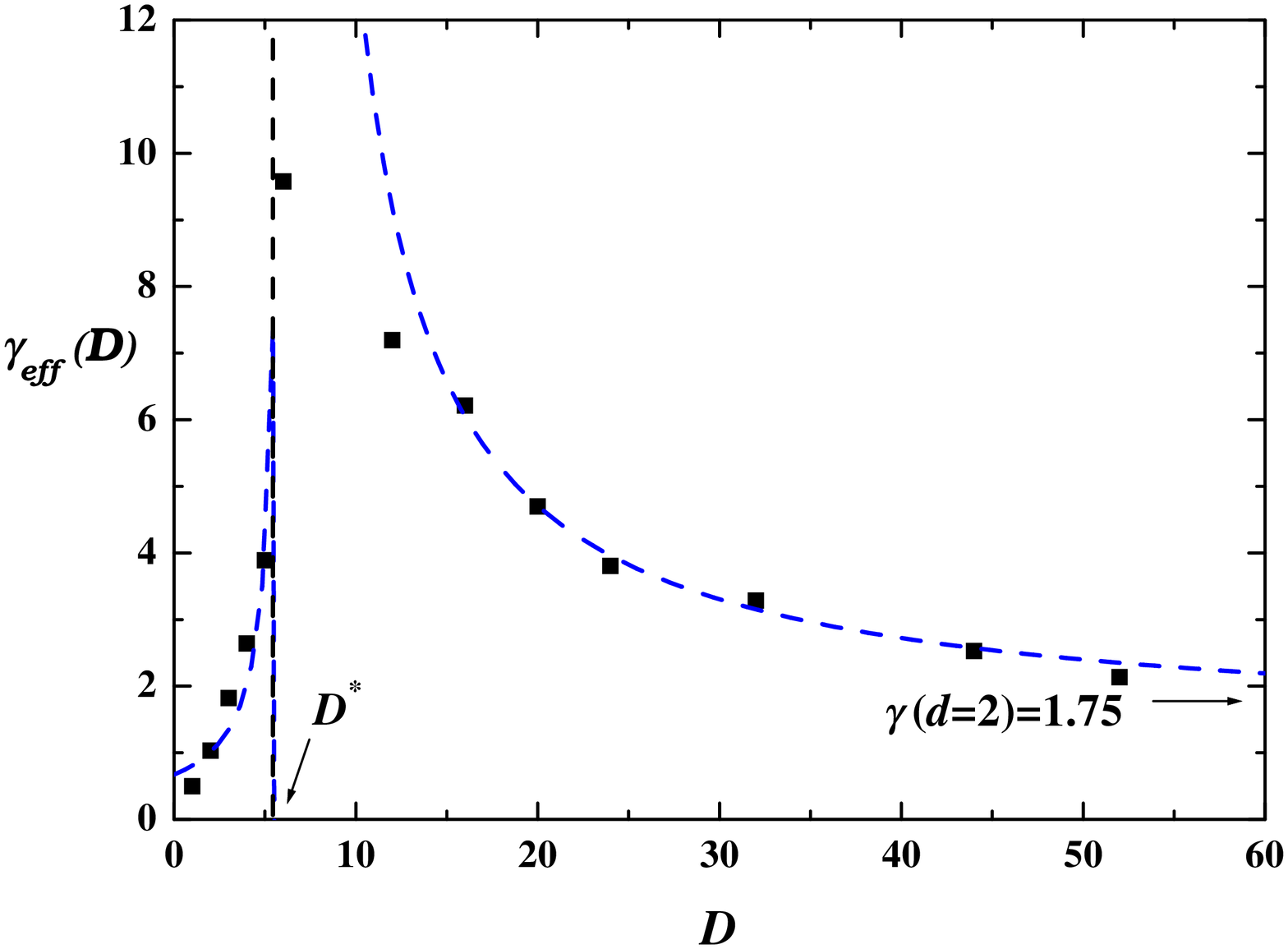}
\\
\caption{\label{fig:Fig5}Plot of susceptibility critical exponent $\protect\gamma_{eff} (D)$ for $%
L\times D$ strips as a function of $D$ for strips with various widths $%
D<D^{\ast }$ and $D>D^{\ast }\simeq 6$. The continuous curve for $D>D^{\ast
} $ is given by $\protect\gamma_{eff} (D)=G+A(D-D^{\ast })^{-1}-B(D-D^{\ast })$
with $G=11/8$, $A=1125/24$ and $B=3/3200$ in good agreemente with $\protect%
\gamma (d)=\protect\beta (d)[\protect\delta (d)-1]$ as determined by from $%
\protect\beta (d)$ and $\protect\delta ^{-1}(d)$ linearly extrapolated from
the respective $d=2,$ $d=4$ values.}

\end{figure}

Figure \ref{fig:Fig5} gives a plot of $\gamma_{eff} (D)$ which shows directly how
this critical exponent blows up as $D$ approaches $D^{\ast }\simeq 6$,
Equation (\ref{eq:4}) gives $\gamma (d)$ in terms of $\beta (d)$ and $\delta
^{-1}(d)$, as specified by Equations (\ref{eq:2}) and (\ref{eq:3}). As shown, 
$\gamma_{eff} (D)$ is given by

\begin{equation}  \label{eq:6}
\gamma_{eff} (D)=G+\frac{A}{D-D^{\ast }}-B\left( D-D^{\ast }\right) \text{, \ \ }%
D>D^{\ast }
\end{equation}

with $G\simeq 11/8$, $A\simeq 1125/24$ and $B\simeq 3/3200$, in good
agreement with $\gamma (d)$ given by Equation (\ref{eq:4}). For $%
D<D^{\ast }$, i.e. for $D=2,3,4,5$, $\gamma_{eff} (D)$ grows up steeply with $D$
and blows up at $D\simeq 6 $, which corresponds to $\beta (d_{c})=0$ where $%
d_{c}=1.333...$, as given by Equation (\ref{eq:2}). $\gamma_{eff} (D)$ for $%
D<D^{\ast }$, i.e. for the interval $1<d<d_{c}$ where $\gamma \simeq \beta
\cdot \delta \simeq \left( 0\right) \cdot \left( \infty \right), $ is
undefined, but it can be fitted reasonably well by

\begin{equation}  \label{eq:7}
\gamma_{eff} (D)\simeq C\cdot \left( D-D^{\ast }\right) ^{-1}\text{, \ \ }%
D<D^{\ast }
\end{equation}

with $C\simeq 4$, which is compatible with $\gamma (d=1)=1/2$.

The hyperscaling relationships \cite{1} result in a critical exponent $\nu $
describing the temperature dependence of the correlation length $\xi $ which
is related to the dimensionality $d$ and to other exponents by

\begin{equation}  \label{eq:8}
\nu^{-1} (d)={y_{1}(d)}=\frac{d}{\beta (d)[\delta (d)-1]}
\end{equation}

For $D\geq D^{\ast }$ the denominator in the right hand side of Equation (%
\ref{eq:8}) can be written as $\{\gamma (d)+2\beta (d)\}\simeq \gamma (d)$
at $d\gtrsim 1.5$. For $D<D^{\ast }$ $(1\leq d\leq 1.333...)$ this
denominator can be approximated by $\{\gamma (d)\}$, because $\beta (d)=0$
for $d\lesssim 1.333...$.Table \ref{tab:table1} gives a set of critical
exponents, including $(\beta /\nu)_{eff} $ and $(1/\nu)_{eff} $, to be used below for
scaling \ Monte Carlo data of $ML^{\beta /\nu }$ vs $|\epsilon |L^{1/\nu }$
for strips of various $D$ and $L=500,1000$ calculated using periodic
boundary conditions.

\begin{figure}
\includegraphics[width=6.1cm,height=7.9cm,angle=270]{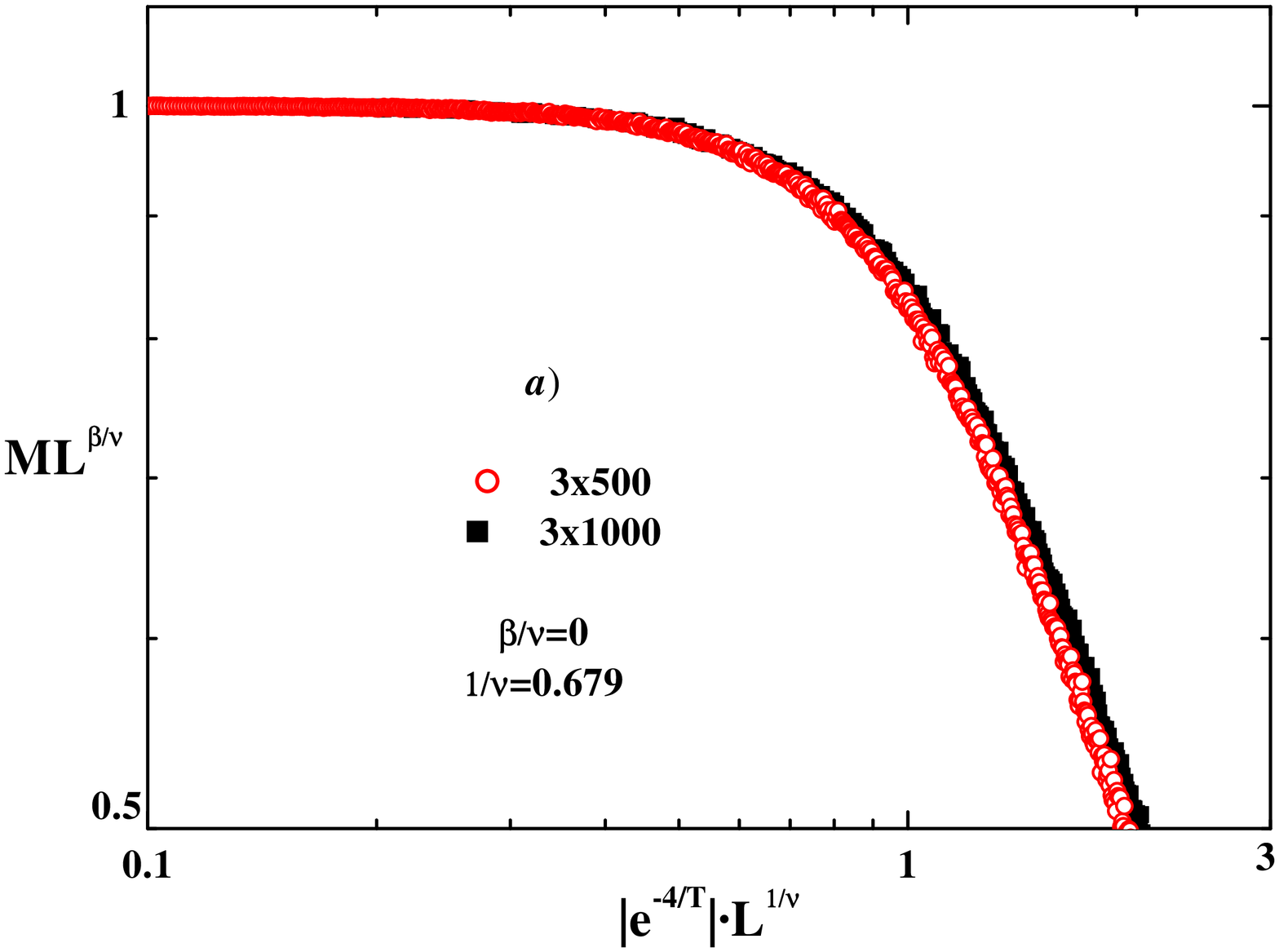}
\\
\caption{\label{fig:Fig6} Scaling plots of $ML^{\protect\beta /\protect\nu }$ vs $|\protect%
\epsilon |L^{1/\protect\nu }$ for strips with $D=3<D^{\ast }$ taking $|%
\protect\epsilon |=|e^{-4/T}|.$}

\end{figure}

\begin{figure}
\includegraphics[width=6.1cm,height=7.9cm,angle=270]{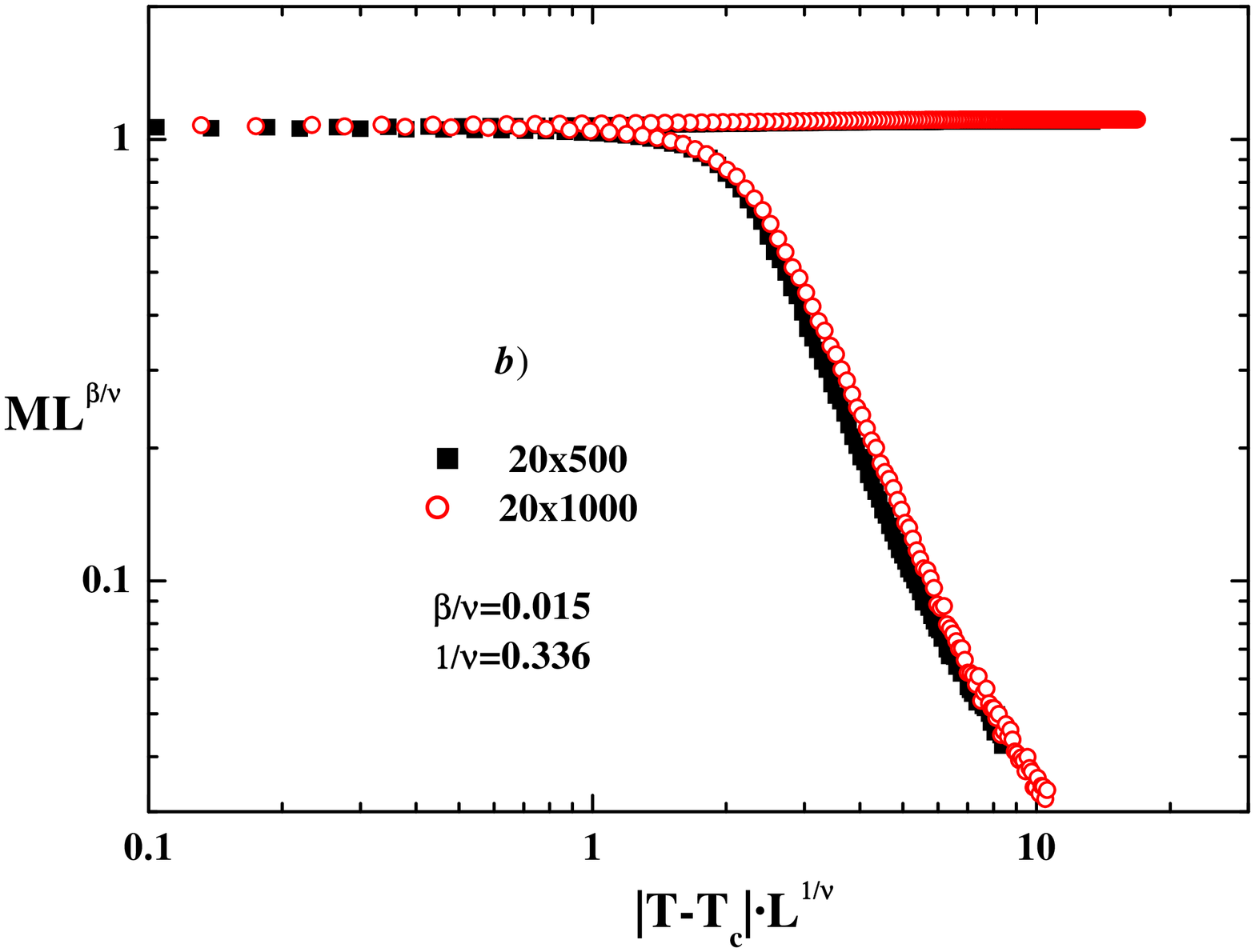}
\\
\caption{\label{fig:Fig7} Scaling plots of $ML^{\protect\beta /%
\protect\nu }$ vs $|\protect\epsilon |L^{1/\protect\nu }$ for strips with $%
D=20>D^{\ast }$ taking $|\protect\epsilon |=|T-T_{c}|.$}

\end{figure}

Figures \ref{fig:Fig6} and \ref{fig:Fig7} give scaling plots of Monte Carlo data for strips of
lenght $L=500,1000$ various widths $D$. Figure \ref{fig:Fig6} shows results for $D=3<D^{\ast }\simeq
6$, taking $|\epsilon |=|e^{-4/T}|$, (one-dimensional scaling) and
Figure \ref{fig:Fig7} results for $D=20>D^{\ast }\simeq 6$, taking $|\epsilon |=|T-T_{c}|$
(two-dimensional scaling).

\section{\label{sec:level1}Concluding remarks}

In summary, our Monte Carlo data properly analyzed show that strips of
dimensions $L\times D$, with $D\ll L$, are{} characterized by a
susceptibility critical exponents $\gamma_{eff} (D)$ blowing up at $D=D^{\ast
}\simeq 6$, which appears to correspond to a dimensionality $d^{\ast }$,
$1.333...\leq d^{\ast }\leq 1.5$, which is well determined by the
extrapolation of $\beta (d)$ and $\delta ^{-1}(d)$ given by the same
linear expressions which describe well the dimensionality dependence for $%
d=2,3,4$. \ This lends empirical support to the theoretical contention that
the renormalization transformations, which are not only scale invariant but
also conformally invariant transformations, imply fractional exponents for
Ising lattices of geometry intermediate between one dimensional and two dimensional.

It may be pointed out that exact information for the largest eigenvalues of the 
tranfer matrix of the system in small width strips (perhaps up to $D\leq 20 $ or so
with infinite length) can be obtained using f.i., the Lanczos algorithm, but, in principle,
this does not result in useful information about the effective exponents, which are
the ones which determine the observable behaviour of $M(T)$ close to, but not arbitrarily
close to the transition.

%
%
\begin{table}
\caption{\label{tab:table1}Ising critical exponents for dimensionalities\newline 1$\leq d\leq2$.}
\begin{ruledtabular}
\begin{tabular}{cccccc}

$d_{eff}$ & $\beta_{eff}$ & $\delta_{eff} ^{-1}$ & $\gamma_{eff}$ & $(1/\nu)_{eff}$ & $(\beta/\nu)_{eff}$\\

\hline

1.0  & 0 & 0 & 1/2 & 2 & 0 \\
1.1  & 0 & 0 & (0.71) & (1.54) & 0\\
1.2  & 0 & 0 & (1.25) & (0.96) & 0\\
1.3  & 0 & 0 & (5.00) & (0.26) & 0\\
1.333...  & 0 & 0 & $\infty$ & 0 & 0\\
1.5  & 0.031 & 0 & $\infty$ & 0 & 0\\
1.6  & 0.050 & 0.013 & 3.700 & 0.421 & 0.021\\
1.7  & 0.069 & 0.027 & 2.516 & 0.642 & 0.044\\
1.8  & 0.087 & 0.040 & 2.100 & 0.791 & 0.069\\
1.9  & 0.106 & 0.053 & 1.887 & 0.905 & 0.096\\

\end{tabular}
\end{ruledtabular}

\footnotetext{The numerical values are obtained by means of the linearly extrapolated
expressions for $\protect\gamma (d)=\protect\beta (d)[\protect\delta (d)-1]$
, $\protect\beta (d)=(\frac{3d}{16}-\frac{1}{4})$, $\protect\delta ^{-1}(d)=(
\frac{2d}{15}-\frac{1}{5})$, $\protect\nu (d)=\protect\gamma (d)/d$ and $
\protect\beta (d)/\protect\nu (d)=(\frac{d}{\protect\delta (d)-1})$. Values
in parenthesis for $D<D^{\ast }$ are approximations obteined, extrapolating
$d_{eff}(D)$ from the $D$ dependence at $D>D^{\ast }$.}

\end{table}

%
%

\begin{acknowledgments}
Support from the DGICyT for grant BFM2000-0032 is gratefully acknowledged.
\end{acknowledgments}

\newpage 
\bibliographystyle{plain}
\bibliography{apssamp}

\newpage

%
%


%

%
%

%
%
%

%
%
%
\newpage
%
%
%
%
%
%
\end{document}